\documentclass[12pt,a4paper,english,fleqn]{article}

\usepackage[sort&compress,round,comma,authoryear]{natbib}
\usepackage[T1]{fontenc}
\usepackage{ae,aecompl}

\usepackage[colorlinks=true,urlcolor=blue,citecolor=blue,linkcolor=red,bookmarks=true]{hyperref}
\usepackage{graphicx}	
\usepackage{float}	
\usepackage{amsmath}	
\usepackage{amssymb}	
\usepackage{xcolor}
\newlength{\lp}
\usepackage[mathscr]{eucal}
\usepackage{amsfonts}
\makeatother
\usepackage{babel}
\makeatother
\begin{document}
\title{The Baryonic Tully-Fisher Relationship: A consequence of Newtonian Gravitation acting in a hierarchical Universe}
\author{D. F. Roscoe (The Open University; D.Roscoe@open.ac.uk)\\ \\ORCID: 0000-0003-3561-7425}
\date{}
\maketitle
\newpage
\begin{abstract}
It has been reported that the application of convolutional neural-network techniques to infer the Dark Matter distribution in the local IGM has revealed how it follows the hierarchical distribution of galaxies in the locality, rather than exhibiting homogeneity.
\\\\ 
This result makes it natural to consider the possibility that, on scales at least as big as $20 \sim 30\,Mpc$, the distribution of \emph{all} material comprising the local IGM is hierarchically distributed. Given this possibility, any model of galaxy formation must then involve a process in which all of the hierarchically distributed material $M_0$ within a sphere $R_0$ coalesces about a unique center so that hierarchical symmetry is broken on the scale $(M_0,R_0)$. 
\\\\
In the particular case of the hierarchical distribution being quasi-fractal $D \approx 2$ in the local cosmos then, for circular velocity $V_0$ on $R_0$, the scaling relation $V_0^4 \sim M_0$ emerges automatically when the condition that such a galaxy formation process must be gravitationally stable in the Newtonian sense is applied. 
\\\\
In other words, subject to the caveat that the analysis applies to a highly idealized model, the Baryonic Tully-Fisher Relationship (BTFR) is shown to arise as a consequence of Newtonian gravitation acting in a hierarchical Universe.
We discuss the ramifications of this result, which are significant and non-trivial.  
\end{abstract}
\newpage{}



\section{Introduction:} \label{Intro}
The modern debate around the idea that material in the local Universe is distributed hierarchically with fractal dimension $D \approx 2$ has been on-going for almost 40 years. Whilst there is now broad agreement that this is the case on \emph{some} scale, the proponents of $\Lambda$CDM cosmology argue for a turn-over to homogeneity on scales of $20 \sim30\,Mpc$, whilst the proponents of the opposite view argue that the hierachical distibution of material persists at least up to $200\,Mpc$. For a fairly recent analysis see  \citet{Tekhanovich} which gives a comprehensive discussion of the issues involved around these divergent opinions. 
\\\\
Irrespective of these divergences, it is a point of considerable significance that the Hubble Law is well established even on scales that are deep inside the smaller of the contested volumes. This very interesting circumstance is the primary  evidence supporting the idea that the IGM is largely populated by an homogeneous distribution of Dark Matter on the small scales required - for, without homogeneity, the linear nature of Hubble's Law cannot be understood within the context of universal expansion.
\\\\
It is for this reason that the paper of \citet{Hong} caused so much consternation: specifically, the authors used state-of-the-art convolutional neural-network techniques combined with modern positional and peculiar velocity data to compute and map the local Dark Matter distribution. Against expectation, this distribution was found to trace the hierarchical distribution of galaxies in the local Universe very closely - there is no indication of homogeneity, and hence no indication that the Hubble Law can be understood in terms of universal expansion.
\\\\
Notwithstanding the Hubble Law problem raised by Hong et al's analysis, the replacement of an homogeneous distribution of DM in the local IGM by a hierarchical distribution of DM makes it entirely natural to consider the possibility that \emph{all} material in the local IGM is distributed hierarchically. In this case, there must be a process of galaxy formation according to which an isolated galactic object can be modelled as a finite bounded  spherically symmetric peturbation of the hierarchical IGM (assumed in the first instance to be an unknown mix of baryonic and non-baryonic mass) - this automatically entails that all of the mass $M_0$ within the sphere $R_0$ has coalesced around a unique centre so that hierarchical symmetry is broken on the scales of $(M_0, R_0)$. The ramifications following from this conclusion are far reaching.
\section{The emergence of the BTFR:} \label{The-model} 
We make the basic assumption that, on local scales at least, the material hierarchy is quasi-fractal $D \approx 2$. Then, for an idealized model:
\begin{enumerate}
	\item The lower cut-off radial and mass scales $(M_0, R_0)$, which define the scales of any given coalesced galaxy, must behave according to
	\begin{equation}
	M_0 = 4 \pi R_0^2 \Sigma_F  \label{(2)}
	\end{equation}
	where $\Sigma_F$ is the mass surface density of the $D=2$ hierarchical mass distribution in the idealized model. In reality, this latter relationship would be a stochastic one, and a rigorous analysis would need to account for that;
	\item Since galaxies in general appear to be stable structures, there must be an equilibrium constraint at the lower cut-off scales of the hierarchy. Using simple Newtonian arguments, we show in appendix \S\ref{SimpleModel} that equilibrium at these lower cut-off scales requires:
	\begin{equation}
	\frac{V_0^2}{R_0} = a_F \equiv 4 \pi G \Sigma_F \label{(3)}
	\end{equation}
	where $V_0$ is the circular velocity on $R_0$ and $a_F$ is the characteristic acceleration scale associated with $\Sigma_F$; 
	\item The relationship
	\begin{equation} 
	V^4_0 = a_F\, G M_0\,,   \label{(4)}
	\end{equation}
	which is formally identical to the Baryonic Tully-Fisher Relationship (BTFR), 
 is now derived directly by eliminating $R_0$ between (\ref{(2)}) and (\ref{(3)}).
	\end{enumerate} 
A rigorous extension of this analysis, beyond the scope of the present note, yields Milgrom's form of the BTFR exactly:
\[
V_{flat}^4 = a_F\, G\, M_{flat} 
\]
with the refinement that $M_{flat}$ is now defined by the scaling relation
\[
M_{flat} \equiv \left(\frac{V_{flat}}{V_0} \right)^2 M_0, 
\]
rather than being empirically estimated.
\section{Empirical support for the BTFR hypothesis} \label{BTFR}
\subsection{The analysis of \citet{Lelli2016B}}
It has only recently been possible to explore the BTFR hypothesis in a statistically rigorous fashion. 
Specifically, the SPARC sample of \citet{Lelli2016A} contains high quality rotation curves and high quality modern surface photometry at $3.6\,\mu m$ for a sample of 175 nearby disk galaxies. The high quality of the surface photometry over this sample allowed \citet{Lelli2016B} to construct photometric models of baryonic mass distributions in that particular subsample of 118 disks which also had rotation curves extending to flatness, making it ideal for a statistically rigorous testing of the BTFR hypothesis.
\\\\
Subsequently, the authors used regression analysis techniques to demonstrate how the subsample really does fit the BTFR (specifically, the quartic power-law form of (\ref{(4)})) with very small scatter. In this way, they argued that the observed scatter is sufficiently below the instrinsic-scatter expectations of $\Lambda$CDM cosmology to present a fundamental difficulty for that cosmology and for the associated idea of dynamically significant quantities of \emph{non-baryonic} matter in the generality of galaxy disks. 
\subsection{The estimation of $(a_F,\,\Sigma_F)$} \label{aFcalc}
\begin{figure}[H]
	\centering
	\includegraphics[width=0.7\linewidth]{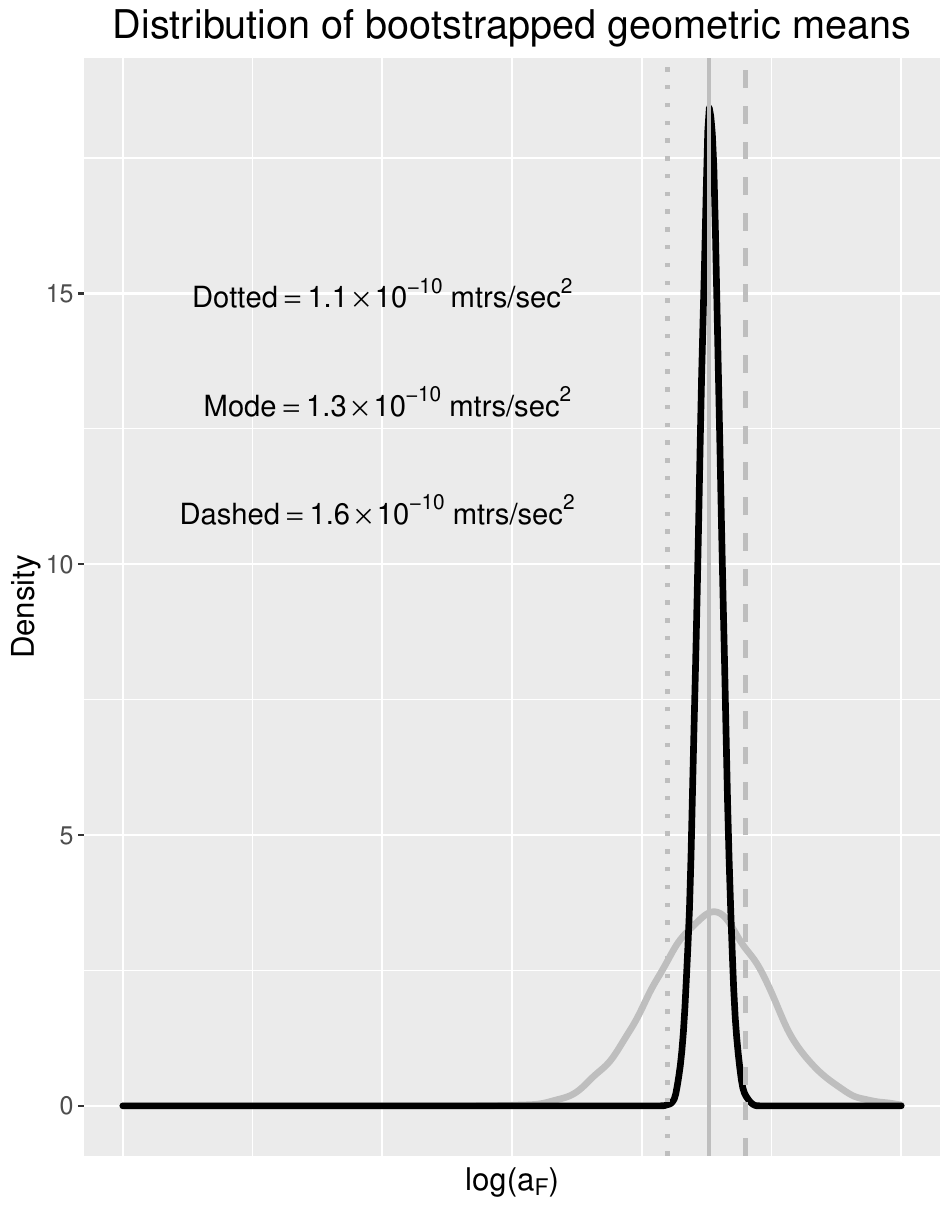}
	\caption{Solid black curve = density distribution of $\log \hat{a}_F$. Solid grey curve arises when velocity and mass data are shuffled with respect to each other. The signal represented by the black curve is destroyed on the shuffled data. }
	\label{GammaDensity2}
\end{figure}

The analysis of Lelli et al was specifically focussed on demonstrating the quartic power-law form of the BTFR. In the following, using the same data, we estimate the value of the acceleration parameter, $a_F$.
\\\\ 
The data used by \citet{Lelli2016B}  is available as an on-line data-sheet giving estimates for the photometrically modelled baryonic masses $M_0$ and flat rotation velocities $V_0$ for the 118 disk galaxies.
Given this data, an alternative demonstration supporting the BTFR hypothesis is provided by showing how the hypothesis, applied differently to the data, yields a very sharp estimate of the characteristic acceleration parameter $a_F$, thereby demonstrating how, for all practical purposes, its value is identical to that of Milgrom's critical acceleration parameter, $a_0$.
\\\\
In order to estimate $a_F\equiv 4\pi G \Sigma_F$ from this data, we rearrange (\ref{(4)}) as
\[
\frac{V_0^4}{G M_0} = a_F \equiv constant
\] 
and hence form the empirical sample distribution
\[
J \equiv \left( \frac{V_{0i}^4}{G M_{0i}}, i = 1... 118 \right).
\]
Then,  from $J$, we generate $N=10000$ bootstrapped distributions, $\hat{J}_k,\,k=1..N$ in the usual way. For each $\hat{J}_k$  we then compute its geometric mean, $\hat{a}_{Fk}$ say,  to obtain, finally, the distribution
\[
A_F \equiv \left(\log \hat{a}_{Fk},\,k=1..N \right).
\] 
The density distribution of $A_F$ is given in figure \ref{GammaDensity2} from which it is clear that the estimate for $a_F$ is very tightly constrained around the modal value
of $1.3\times 10^{-10}\,mtrs/sec^2$ which, for all practical purposes, is identical to Milgrom's value of MOND's critical acceleration parameter $a_0$. This estimate of $a_F$ corresponds to $\Sigma_F \approx 0.15\,kg/mtr^2$ for the mass surface density of the hierarchical cosmos.
\\\\
The grey curve in figure \ref{GammaDensity2} arises from the same analysis but applied to shuffled velocity and mass data. It is clear that the signal so powerfully present in the unshuffled data is destroyed by shuffling. We conclude that, for all practical purposes the signal for the mass surface density  $\Sigma_F \approx 0.15\,kg/mtr^2$ in the hierarchical cosmos is real. 
\subsection{Implications for the hierarchical IGM of  \citet{Hong}}
According to \cite{Hong}, the material of the local IGM consists of undetected matter (their Dark Matter) distributed in the same hierarchical manner as the galaxies in the local Universe. It follows that since (\ref{(4)}) is derived from the hypothesis that galaxies form by coalescing in a stable way out of the $D=2$ hierarchical IGM, then galaxies themselves must consist of the same material that is undetected in the IGM according to Hong et al. But this is the material that is implicated in the BTFR of (\ref{(4)}) which the analysis of \cite{Lelli2016B} has shown to consist purely of baryonic matter.
\\\\
In other words, the hierarchical distribution of undetected material which comprises the local IGM according to the analysis of Hong et al must consist of baryonic matter according to the analysis of Lelli et al. This raises a question about the nature of the undetected baryonic matter within the IGM. We consider this question in \S\ref{DarkMatter}
\section{The nature of undetected baryonic matter?} \label{DarkMatter}
It has been argued that the Dark Matter of the IGM is distributed in a $D=2$ hierarchy and consists of undetected baryonic material. So, the immediate question is: \emph{how can such a distribution of baryonic material remain undetected?} There are two potential strands to the answer, easily conceived to be acting in concert.
\subsection{Conventional possibilities}
Given that the local IGM forms a $D=2$ hierarchy, then its volume density in a spherical volume of radius $R$ tends to zero as $R \rightarrow \infty$, making its detection at large radii intrinsically difficult.
\\\\
Furthermore, it can reasonably be assumed that the IGM is at least close to being in thermal equilibrium with the general background, again making its detection against the background also intrinsically difficult.
\subsection{Unconventional possibilities} \label{BlackBody}
Until recently, it has always been assumed that there is no such thing in nature as a perfect, or near perfect, blackbody absorber - the reason being that no such thing had ever been observed. However, \citet{Mizuno} showed how to fabricate, from agglomerations of single-walled carbon nanotubes (SWCNTs), material distributions  having specific bulk statistics which act as near-perfect blackbody absorbers (emissivity $> 0.98$) across a very wide range of incident wavelengths from UV at 200$nm$ to the far IR at 200$\mu m$.  
\\\\
This behaviour has been shown to be independent of the specific properties of the individual SWCNTs, but is rather a consequence of the bulk statistical characteristics of the fabricated SWCNT distributions. We know that many allotropes of carbon exist in interstellar space and these must to some extent be blown into the IGM from the generality of galactic interiors. It is a short step to visualizing the existence of clouds of SWCNTs dispersed throughout the hierachical IGM  containing sub-populations which,  when viewed in projection along any given line of sight, possess the bulk statistical characteristics required to mimic the properties of the fabricated SWCNT distributions of \citet{Mizuno}. 
\\\\
In this way, it is possible to conceive how SWCNT clouds within the IGM have the potential to act as \lq{dispersed near-perfect blackbody objects}' making them virtually undetectable. 
\section{Summary and conclusions:}
The Baryonic Tully-Fisher Relation (BTFR) is a purely empirical scaling relation which the work of \cite{Lelli2016B} has shown to be well supported on modern data. But any understanding of its basis in fundamental physics has remained elusive. 
\\\\
Given that Milgrom's MOND can be interpreted primarily as an empirical mechanism by which this scaling relationship is imposed to constrain the dynamics of certain astrophysical systems, then MOND can never be understood until the origins of the BTFR are themselves understood. To this end, in allowing us to conceive that the distribution of \emph{all} material in the IGM follows the hierarchical distribution of galaxies in the local cosmos,  the results of \cite{Hong} have provided an unexpected pathway to a qualitative understanding of the BTFR.
\\\\
In particular, given that the IGM hierarchy is quasi-fractal $D\approx 2$ on some scale, we necessarily have a galaxy formation process in which all of the matter $M_0$ in a sphere $R_0$ coalesces about a unique center so that hierachical symmetry is broken, with $(M_0,R_0)$ then representing the lower cut-off scales of the hierarchy. If the galaxies so formed by this process are constrained to be gravitationally stable in the Newtonian sense, then the relationship $V_0^4 = a_F G M_0$ arises automatically. Whilst this is identical to the Milgrom form of the BTFR only for the special case $V_0=V_{flat}$, the exact Milgrom form is recovered via a rigorous extension of the current analysis which is beyond the scope of the current note.
\\\\
As a direct consequence of these considerations, it is found that the IGM must then be constituted primarily of (undetected) baryonic material which raises the obvious question: \emph{why is it so difficult to detect?} Work in the material sciences \citet{Mizuno} raises the possibility that the IGM contains large clouds of single-walled carbon nanotubes blown from the generality of galaxy interiors, the bulk statistical properties of which allow these clouds to act as near-perfect blackbody absorbers. Given enough of this material, the hierarchical IGM would be extremely difficult to detect. 
\subsection*{Data availablity statement}
The data underlying this article and used in \S\ref{BTFR} are publically available to download via  \citet{Lelli2016B} from the ApJL website.

\appendix
\section{Equilibrium at the lower cut-off scales of the hierarchy}
\label{SimpleModel}
In the cosmos of our experience, galaxies in general appear to be stable and long-lasting structures. Since the matter distribution in the $D=2$ fractal hierarchy is isotropic (by definition) about any arbitrarily chosen centre, then the notional gravitational acceleration imparted to a particle at radius $R$ from the centre, and generated by the material contained \emph{within} $R$, is directed towards the chosen centre and has magnitude given by
\begin{equation}
\frac{M(R) \, G}{R^2} = 4 \pi G\, \Sigma_F \equiv a_F,\,\,\, R< \infty. \label{aF}
\end{equation}
On this basis, it is clear that the \emph{net} actual gravitational acceleration imparted to a material particle immersed anywhere in the global hierarchy is zero, from which it can be concluded that a $D=2$ fractal distribution of material is in a state of dynamical equilibrium.
\\\\
It follows that:
\begin{itemize}
	\item if a finite spherical volume, radius $R_0$, is imagined emptied of all material, then the net actual gravitational acceleration of any material particle placed on $R_0$  will be $a_F$ directed radially outwards from the centre of the empty volume;
	\item the empty spherical volume is unstable since all accelerations on $R_0$ are outward. It follows that stability requires the volume to be occupied by a stablizing mass, a galaxy say, creating a state of zero net radial acceleration on $R_0$. In other words, the equilibrium condition
	\begin{equation}
	g_0 \equiv \frac{V_0^2}{R_0} = a_F \label{eqn3}
	\end{equation}
	must be satisfied.
\end{itemize}



\end{document}